
\magnification=\magstep1
\hsize 6.0 true in
\vsize 9.0 true in

\voffset = - .2 true in

\font\tentworm=cmr10 scaled \magstep2
\font\tentwobf=cmbx10 scaled \magstep2

\font\tenonerm=cmr10 scaled \magstep1
\font\tenonebf=cmbx10 scaled \magstep1

\font\eightrm=cmr8
\font\eightit=cmti8
\font\eightbf=cmbx8
\font\eightsl=cmsl8
\font\sevensy=cmsy7
\font\sevenm=cmmi7

\font\twelverm=cmr12
\font\twelvebf=cmbx12
\def\subsection #1\par{\noindent {\bf #1} \noindent \rm}

\def\mid {\let\rm=\tenonerm \let\bf=\tenonebf \rm \bf}

\def\para{\par \vskip 12 pt}

\def\head{\let\rm=\tentworm \let\bf=\tentwobf \rm \bf}

\def\heading #1 #2\par{\centerline {\head #1} \smallskip
 \centerline {\head #2} \vskip .15 pt \rm}

\def\eight{\let\rm=\eightrm \let\it=\eightit \let\bf=\eightbf
\let\sl=\eightsl \let\sy=\sevensy \let\m=\sevenm \rm}

\def\foots{\noindent \eight \baselineskip=10 true pt \noindent \rm}
\def\sexion{\let\rm=\twelverm \let\bf=\twelvebf \rm \bf}

\def\section #1 #2\par{\vskip .35true in \noindent {\mid #1} \enspace {\mid #2}
  \para \noindent \rm}

\def\abstract#1\par{\para \foots {\bf Abstract: \enspace}#1 \para}

\def\author#1\par{\centerline {#1} \vskip 0.1 true in \rm}

\def\abstract#1\par{\noindent {\bf Abstract: }#1 \vskip 0.5 true in \rm}

\def\midsection #1\par{\noindent {\sexion #1} \noindent \rm}

\def\sqr#1#2{{\vcenter{\vbox{\hrule height.#2pt
  \hbox {\vrule width.#2pt height#1pt \kern#1pt
  \vrule width.#2pt}
  \hrule height.#2pt}}}}

\voffset=-.5truein
\vsize=9truein
\baselineskip=16pt
\hsize=6.0truein
\pageno=1
\pretolerance=10000
\def\n{\noindent}
\def\s{\smallskip}
\def\b{\bigskip}
\def\m{\medskip}
\def\c{\centerline}

\centerline{\bf\mid INHOMOGENEOUS CHAOTIC INFLATION}
\b
\n {S. Mukherjee$^*$,
Physics Department, Sharif University of Technology,
Tehran, Iran}
\s
\centerline {and }
\s
\n {N. Dadhich$^{**}$,
Inter-University Centre for Astronomy \&\ Astrophysics,
P.B.No.4, Pune 411 007, India}
\b
\b
\n {\bf Abstract}
\s
\baselineskip=24pt
\n     A chaotic model of the early universe within the framework of the
singularity-free solutions of Einstein's equation is suggested. The evolution
of our universe at its early stage, starting out as a small domain of the
parent universe, is governed by the dynamics of a classical scalar field $\phi$
. If in any such domain, larger than
Planck length,$\dot \phi$ happens to be very large,$\phi$ may develop a
dominant
inhomogeneous mode,leading to an anisotropic inflation of the universe.
The particle $\phi$ is coupled to other particles, which are produced copiously
after inflation and these thermalize leading to a rather low temperature
universe $(T \geq 10^{4} $ Gev). The electroweak  B+L Baryogenesis is assumed
to
account for the observed baryon asymmetry. The universe now passes through
a radiation-dominated phase, leading eventually to a matter-dominated
universe, which is isotropic and homogeneous. The model does not depend on
the details of Planck scale physics.
\b
\b
\b
\n$^{*} $Permanent address : Physics Department, North Bengal University, Dist.
Darjeeling, India 734 430. (E-mail address : sailom@nbu.ernet.in )
\s
\n$^{**} $E-mail address : naresh@iucaa.ernet.in

\vfill\eject
\baselineskip=24pt
\n In the chaotic inflationary models, the universe is assumed to evolve out
of a random distribution of initial data. This  makes the model superior
to other inflationary models, which require a very special class of potentials
and/or specific initial conditions. For example, in the new inflationary model,
the universe is assumed to be in thermodynamical equilibrium initially, with
the inflaton field $\phi$ at the  local minimum of its temperature - dependent
potential V($\phi$,T). On the other hand, the chaotic model should work, in
principle, for a random distribution of the initial values of $\phi$ and for a
large class of potentials. However, as Linde $^{1} $ has pointed out, the
values of $\phi $
are not totally unspecified. When one specifies the initial conditions for a
classical space-time, one must have
$$\rho < M^4_p, ~ V(\phi)<M^4_p,~(\partial_i \phi)^2 < M^4_p, ~ i~ = ~0,..,3
\eqno (1) $$

\s
\n These constraints make the scalar field $\phi$ almost homogeneous, if it is
large,
as can be seen from the following considerations. A closed mini-universe,
emerging out of the quantum space-time foam, has a typical size of Planck
length, $M^{-1}_p $. In such a universe, with the gradient being less than
$M^{2} _p $,the field $\phi$ cannot differ from some average value $\phi_0 $
by more than  $ M_{p}.$
Thus, if the value of $\phi_{0}  \gg M_{p}, \phi $ is essentially homogeneous
to begin with.
Linde has assumed a similar homogeneous and large value for $\phi$ even for an
open universe. However, the fact that $\phi \gg M_{p} $ leads to a very low and
`unnatural' value for the coupling constants (e.g. $ \lambda \sim 10^{-13}  $
for a
$ \lambda \phi^{4}  $  theory). Moreover, the situation with the open universe
is not
very clear. It would indeed be useful if a scenario with $\phi_0 < M_{p }$ also
works.
Obviously, one has to encounter an anisotropic and inhomogeneous space-time
and also an inhomogeneous $\phi $. But such a model will be more `natural' and
also
in keeping with the expectation that the universe would have evolved to its
present state even if the initial geometry is not carefully chosen. A first
step in this direction will be  to consider a mini-universe with lesser
symmetry (say one with cylindrical symmetry) and try to build a chaotic
model with initial values of $\phi $ less than $M_{p }$. We present here such a
model,
which differs considerably from the model of Linde. Also, in our model, the
universe need not pass through states of very high temperature or density.
The universe remains inhomogeneous and anisotropic  till very late (till
matter-
dominated era). The inflation takes place in two stages. Our description
of space-time makes use of the singularity-free solutions$^{2,3} $. These
solutions describe an inhomogeneous and anisotropic space-
time, containing a perfect fluid. These may be generalized to describe
different
stages of the evolution. Ruiz and Senovilla$^{3} $ have given a useful
factorizable
expression for the metric, describing a space-time, which admits two commuting
space-like Killing vector  fields,  which are mutually as well as hypersurface
orthogonal.
We follow their notations but consider only a subfamily, viz.
\s
$$ ds^{2} = T^{n+1} ~F^{2} (-dt^{2} +
H^{2}dr^{2}) ~+ ~T^{1+n} G P d\theta^{2} ~+ ~T^{1-n } GP^{-1} dz^{2}  \eqno (2)
$$
\s
\n where $ T\equiv T(t) $ satisfies  the equation
\s
 $$ { \ddot T \over T}   = \epsilon a^{2}, \epsilon = 0, \pm 1,  \eqno (3) $$
\s
\n with $a $ an integration constant.  The functions F, H, G, P are functions
of r only. These are related by
\s
$$F^{2} = (P/G)^{n} \eqno (4) $$

\s
\n  and H, G and P satisfy the differential equations :
\s
$$ \alpha^{\prime} ~+~ \alpha\beta ~-~ \alpha ~ {H' \over H }~ = ~\epsilon n
a^2  H^2 \eqno (5) $$
\s
$$ (1-n) \beta^{\prime} + n \beta^2 - 2n \beta\alpha + \alpha^2 - (1-n)~\beta ~
{H^{\prime}
\over H }= n(1-n) \epsilon  a^2 H^2 \eqno (6) $$
\s

\n with $ \alpha = {P^{\prime } \over P } $  and  $ \beta = {G^{\prime } \over
G }$   . In above, dot and prime denote  differentiation w.r.t.
t and r respectively. Since there are only three relations among the four
functions
F,H,G,P,   they will be given in terms of an arbitrary function $C(r) $.
However, we shall
impose, when convenient, an additional relation, thus determining these
functions completely.
\s
\n Some interesting features of these non-singular solutions have already been
 studied$^{4-6} $.  The singularity free family (2) can be considered as
arising out of FRW metric for open universe through a natural inhomogenisation
and isotropisation process$^{4} $ and it is unique$^{5} $ and geodesically
complete$^{6} $. If one wishes to study the cosmology of a closed universe,
evolving out of the quantum foam, the non-singular nature of the solution
is of no special advantage. However, we wish to study open universes as well.
These non-singular solutions may be useful, if one tries to give a classical
description of an ever-existing parent universe. We will consider this
problem elsewhere. For the purpose of the  model, it is sufficient to assume
that there is an ever-existing parent universe and to follow the evolution of a
small
domain (larger than Planck size) of this universe, which happens to satisfy
suitable initial conditions, to be given later on. The evolution of this
mini-universe is governed by the dynamics of a scalar field $\phi$, having an
effective potential V($\phi$). It can be shown that the assumed cylindrical
symmetry of the space-time ensures that $\phi $ will not depend on $\theta $
and z,
i.e.$\phi \equiv \phi(t,r) $. The relevant field equation, with the metric (2),
is given by
$$ -\ddot \phi  - \dot\phi {\dot T \over T} + {\phi^{\prime \prime} \over H^2}
+
{\phi^{\prime} \over H^2} {(GH^{-1})^{\prime} \over (GH^{-1})} = {dV(\phi)
\over d \phi }
T^{n+1} F^2 . \eqno (7) $$
\s
\n Note that the equation gets simplified, if we impose the condition G = H. We
shall follow the evolution in steps and outline the relevant features :
\b
\n 1.  {\it  First inflationary stage :}
\b
\n      In this model, in order to qualify as a mini-universe, a small domain
of space-time has to satisfy the following conditions :
\s
\n Initially , $ \dot \phi$ should be much larger than $\phi$ as well as
$\phi^{\prime }
(\phi^{\prime \prime} $if  $H = G),$ so that the relevant equation of state is
that of
stiff matter.
\s
\n When these conditions are met, the mini-universe can have a metric of the
form (2), for $n < -1 $. With G = H, we have a class of solutions
\s
$$ C(r) = \lbrack -na^2M (r+q) \rbrack^{-n}, $$
\s
$$ P^2(r) = n^2a^2M^2C^2(r), ~ H(r) = G(r) = {1 \over a(r+q)},$$
\s
$$ T(t) = T_0 Cosh (at), ~\chi p = \chi \rho = {1 \over 2} \dot \phi^2
T^{-(n+1)} ~C^{1-n} \eqno (8) $$

\s

\n where M, q and $a $ are constants. We note that the above solution leads
to an inhomogeneous expansion given by
\s
$$ V(t) = V(0) exp \lbrack - {(n+3) \over (n+1)F(r)} \cdot T(t)^{-{1 \over 2}
(n+1)} \rbrack  \eqno(9) $$
\s

\n where V(0)  is the volume of the miniuniverse at t = 0. The solution with
$n = -3 $  may be useful to describe an ever-existing classical space-time or
a part thereof, as it permits no expansion. While any solution with $-3>n>-1 $
is relevant, we consider, in particular, the solution with $n = -2 $. It gives
an inflationary universe which expands as
$$ V(t) = V(0) exp \lbrack {1 \over F(r)} \sqrt {cosh(at)} \rbrack .\eqno (10)
$$
\s

\n The function F(r) in this case is given by
\s
$$ F(r) = {1 \over 8 a^6 M^3  (r+q)^3}  \eqno (11) $$
\s
\n so that a suitable choice for q and $a $ can provide adequate inflation
everywhere. The space-time is
highly anisotropic, there being a rapid expansion along the z-axis and
contractions along directions normal to it. The inflation ceases because
the field $\phi$ increases rapidly. To see this, we rewrite equation (7),
neglecting the $\phi^{\prime \prime} $ term, as
\s
$$ \ddot \Psi + \dot \Psi {\dot T \over T} = - 4 \lambda \Psi^3 T^{n+1}, \eqno
(12) $$
\s

\n where we have substituted V($\phi) =  \lambda\phi^{4}  $ and $ \Psi = \phi
F(r) $.
Since $\phi$ is small initially, we neglect the R.H.S. and obtain a
solution for the increasing mode,
\s
$$ \Psi = \Psi_0 + {2A \over a} tan^{-1} (e^{at}) - {\pi A \over 2a} \eqno (13)
$$
\s

\n where  $\Psi_0 $      is the value at t = 0. We choose A as a constant and
not a
function of r to restrict the discussion to the case of simple inhomogeneity.
Note that $ \Psi $     approaches rapidly to a limiting value, $ \Psi_1 =
\Psi_0 + {A\pi \over 2a} $
and if  ${\pi A \over 2a} {\gg \Psi_0 } $                    everywhere in the
miniuniverse, we have almost
a homogeneous $ \Psi $   . However, this should happen before the potential
term
becomes important and the stiff matter equation of state gets appreciabely
disturbed. This gives the condition
\s
$$ \dot \Psi \dot T \gg 4 \lambda \Psi^3 $$
\s
\n or
\s
$$ \lambda \ll {2a^4 \over \pi^3 A^3} .\eqno (14) $$
\s

\n We, however, want $ {\pi A \over 2a} $    to be much smaller than $M_p $.
The singularity free  class of solutions
seem to be very rich and can possibly accommodate the case when the potential
term makes a contribution to the energy density and pressure. The class also
includes a solution relevant  for a cosmological constant  $ (m = {1 \over 2},
n=-1) $
but the inhomogeneous  model does not pass through this stage. Thus, the
usual potential-driven inflation does not occur in this model. As the field
$\phi$ increases, there is a change in scenario, because $\phi$ is presumed to
be
unstable, being coupled to other fields, both    scalars  and fermions.
Eventually,
$\phi$ will decay producing these particles, which thermalize. It is known$^{1}
$
that one can get a hierarchy of temperatures depending on how $\phi$ decays.
The mechanism of baryogenesis gives a lower bound on the temperature. If
baryogenesis is due to B + L  violating electroweak processes, a ` reheating'
temperature $ \sim 10^{4} $ Gev is adequate. There does not seem  to be any
compelling reason for the universe to pass through a stage of higher
temperatures.
After thermalization, the universe is full of radiation and another
inflationary
stage begins.
\b
\n 2.  {\it   Second Inflationary stage :}
\b
\n A radiation dominated universe can be described by the metric (2) for n = 3.
This
is the original Senovilla solution. We have indeed a large class of solutions :
\s
$$ ds^2 = T^4 C^2 (-dt^2 + H^2dr^2) +T^4 GP d \theta^2 + T^{-2} GP^{-1} dz^2
\eqno (15) $$
\s

with  $ T = T_0 cosh (at), C(r)$ an arbitrary function,
\s
$$ \rho = 3p =  {5 \over 3} k ~T^{-4}C^{-4} \eqno (16) $$
\s

\n with  $ k \not= 0. $  There is now a contraction along the z- direction and
expansions
along directions normal to it. This peculiar two stage inflation helps to make
the universe isotropic. The expansion during this stage can be calculated from
the
relation,
\s
$$ V(t_2 ) = V(t_1) exp~\bigg \lbrack {3 \over 2F} \bigg({1 \over
Cosh^{2}at_{1}} - {1 \over Cosh^{2}at_{2}}\bigg)\bigg \rbrack \eqno (17) $$
\s

\n with  $ t_2 > t_1. $  Since both energy density and pressure fall as
  $ T^{-4}  $   ,
it is obvious that massive particles will soon become non-relativistic and a
matter-dominated stage will be ushered in.
\b
\n 3. {\it  Matter - dominated era :}
\b
\n After the second inflation, we expect the universe to become isotropic and
homogeneous and go over to the FRW universe. Actually, within the framework
of the metric (2), there is only one window through which this can happen
and this is during the matter-dominated era. The solution is
\s
$$ n = 0, F(r) = H(r) = 1, G(r) = P(r) = r  \eqno (18) $$
\s
\n corresponding to the metric
\s
$$ ds^2 = T(t) \lbrack -dt^2 + dr^2 + r^2 d\theta^2 + dz^2 \rbrack. \eqno (19)
$$
\s

\n Einstein's equations now reduce to
\s
$$ G_{00} = T^{-1} \cdot {3 \over 4} {\dot T^2 \over T^2} = \chi \rho
\eqno (20) $$
\s
$$ G_{11} = G_{22} = G_{33} = T^{-1}  \lbrack - {\ddot T \over T} + {3 \over 4}
{\dot T^2 \over T^2 } \rbrack = 0  \eqno (21) $$
\s
\n which has the solution,
\s
$$ T(t) = Bt^4,  \rho (t) = {12 \over B t^6} .\eqno (22) $$

\s
\n We can rewrite the results in the conventional form by introducing a new
time  $\eta $  , viz.

$$ ds^2 = -d \eta^2 + a^2(\eta) \lbrack dr^2 + r^2 d\theta^2 +dz^2 \rbrack
\eqno (23) $$
which is the well-known Einstein-deSitter universe. The model thus leads to a
flat space with $ \rho \sim a^{-3}, a \sim \eta ^{{2 \over 3}} $
          . It explains in a natural
way why the universe is so flat or why $ \rho $   is so close to the FRW
critical density.
We have introduced a new time  $ \eta $ , but this is the only time of physical
interest, because no astrophysical phenomena of  earlier epochs are
observable to us.
\s
\n It is useful to show that the universe does tend to roll down to the
symmetric
state given by equation (19). A simple way to do this is to study the stability
of the solution (19). We put H = 1, which can be done without any loss of
generality. If we now consider small linear perturbations of the functions
P($r$) and G($r$) about the solution (19), making use of the field equations
(5)
and (6), the solution is seen to remain invariant, modulo a trivial scaling.
True
to our expectation, the symmetric state is most stable, and given sufficient
time, the universe will roll into it.
\s
\n To summarize, we have presented a scenario involving a set of initial data
not considered in Linde's chaotic model. In a realistic chaotic model of the
early universe, one should consider the entire phase-space and also include the
corrections due to quantum fluctuations of the scalar field. Although our model
is based on a particular class of inhomogeneous solutions, it is quite possible
that initial inhomogeneities of many different types may also permit evolutions
leading eventually to a large homogeneous and isotropic universe. An
understanding of all these possibilities will be useful to arrive at the Grand
chaotic model. The effect of quantum fluctuations of $\phi $   is presently
under study and will be reported elsewhere.
\b
\n S.Mukherjee would like to thank Inter-University Centre for Astronomy and
Astrophysics,
Pune and the Physics Department, Sharif University of Technology, Tehran for
hospitality during the period of this work and Reza Mansouri for useful
discussions.
\vfill\eject
\c {\mid References :}
\item{1.}  A.D. Linde, Inflation and Quantum Cosmology (Academic Press, NY
1990).
\s
\item{2.}  J.M.M. Senovilla, Phys. Rev. Lett. {\bf 64}, 2219 (1990).
\s
\item{3.}  E. Ruiz and J.M.M. Senovilla, Phys. Rev. D{\bf 45}, 1995 (1992).
\s
\item{4.}  N. Dadhich, R. Tikekar and  L.K. Patel , IUCAA - 16/93 (Preprint).
\s
\item{5.}  N. Dadhich, L.K. Patel and R. Tikekar, IUCAA - 20/93 (Preprint).
\s
\item{6.}  N. Dadhich and L.K. Patel, IUCAA - 21/93 (Preprint).

\bye